# A study on transmitted intensity of disturbance for air-spaced Glan-type polarizing prisms

Ji-Yang Fan [a], Hong-Xia Li [b], Fu-Quan Wu [b]

[a] *Department of Physics, Qufu Normal University, Qufu 273165, P.R. China*
[b] *Institute of Laser, Qufu Normal University, Qufu 273165, P. R. China*

**Abstract**

We study theoretically and experimentally the transmission intensity of polarized light through the air-spaced Glan-type polarizing prsims. It is found that the variation of the transmitted intensity with the rotation angle deviates from Malus Law, exhibiting a cyclic fluctuation with the rotation angle. The occurrence of the disturbance is explained by the use of an argument based on the interference effect produced from the air-gap in the prisms. The theoretcal results are well agreed with the experimental ones. By selecting the proper cut angle of prism and reducing the thickness of air-gap in prism, the disturbance may be minimized.

## 1.Introduction

Polarizing prisms are very important passive devices in optical information processing, laser modulations and laser measurements, etc. Air-spaced Glan-type polarizing prisms are among the most common types of prisms in use today [1-3]. They are made out of optical calcite, which exhibits strong birefringence over a wide wavelength range. They can be used at shorter wavelengths than cemented prisms with the spectral transmission range being from the ultraviolet to the near infrared (about $300\sim 2500$ nm). The extinction ratio of the Glan-type prisms is better than $10^{-5}$ [4], and the light is nearly uniformly polarized over the field. There is no lateral displacement in the apparent position of an axial object viewed through a Glan-type prism. Moreover they show a good quality of avoiding overheating when high-powered lasers are used [5].

In theory the transmitted intensity should follow the Malus cosine-squared law when polarized light is normally incident on an air spaced Glan-type polarizing prism. In experiment, however, we found that the variation of the transmitted intensity with the rotation angle deviates from Malus Law, exhibiting a cyclic fluctuation with the rotation angle. The effect is unfavorable for the quality of the transmitted polarized light. Therefore, it depresses the precision of measurement as the polarized prism is used in light path. The disturbance for the Glan-Foucault prism is even much stronger than that for the Glan-Taylor prism. In this paper, we report the experimental results of the disturbance in transmitted intensity with the rotation angle. By the analysis to the above results, we find that the interference induced by air-gap in Glan-type prism lead to the disturbance. A theoretical analysis gives that optimizing the cut angle of prism and controlling the thickness of the air-spaced layer may minimize the disturbance, which are surely of guiding significance for designing high performance optical systems and for choosing suitable polarized prisms for different optical experiments.

## 2. Experiments and results

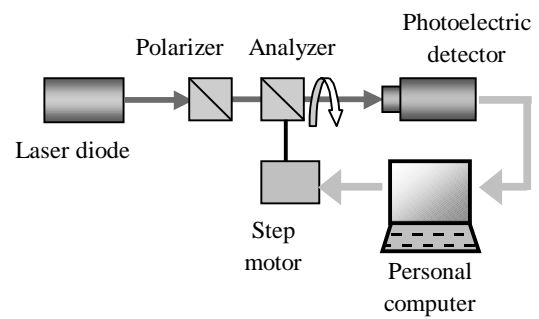

Fig.1. Experimental setup for measuring the transmitted intensity.

The experimental setup is shown in Fig.1. The monochromatic light with the wavelength of 650 nm will be linearly polarized with an intensity of $I_0$ after it passes through a polarizer. Then it normally incidents to an air-spaced Glan-type polarizing prism (the analyzer) fixed on the stepmotor that can rotate continuously around the prism axis, so that the angle between the direction of the optic axis of the polarizer and that of the analyzer can be changed successively. The transmitted intensity is detected by the photoelectric detector and then put into a computer, and processed by the testing software of polarizing light intensity. The measured results of the transmitted intensity versus the rotation angle are directly displayed on the screen. Signals sent out from the computer can control the rotation of the stepmotor. According to Malus cosine-squared law, the transmitted intensity of rays coming through the air-spaced Glan-type polarizing prism is obtained as

$$I = I_0 \cos^2 \phi, \tag{1}$$

where $\phi$ is the angle between the optic axis of the polarizer and that of the analyzer, that is the rotation angle. It is found in the experiment that the transmitted intensity approximately obeyed the Malus law except for the appearance of the disturbance, which is most obvious while the two optic axes are parallel to each other. The experimental curves for the transmitted intensity in one rotation period for the Glan-Taylor prism and for the Glan-Foucault prism are given in Fig.2 (a) and (b), respectively, which shows a stronger disturbance in the Glan-Foucault prism than in the Glan-Taylor prism.

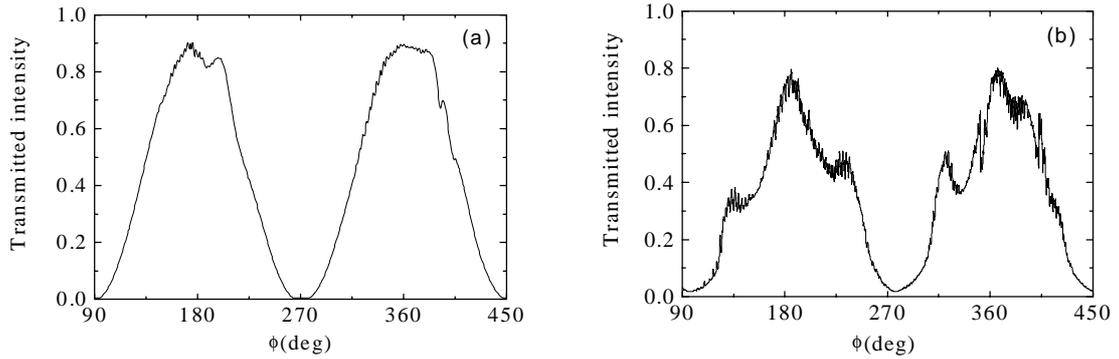

Fig.2.   Experimental curves of transmitted intensity for (a) Glan−Taylor prism and for (b) Glan−Foucault prism.

## 3. Theoretical Analysis and Numerical Results

In this part we shall present a theoretical explanation for it. An air-spaced Glan-type polarizing prism is composed of two right-angle Iceland crystals [1]. The monochromatic light traveling in the front half of the prism is broken up into the ordinary ray $o$ and the extraordinary ray $e$, while still propagating in the same direction. When they are incident upon interface between the crystal and the air-gap, the $e$ ray passes through while the $o$ ray is totally reflected. The $e$ ray will be reflected for several times before it is transmitted through the air-gap. The reflected $e$ ray in the air-gap can interfere with $e$ ray that directly passes through into the air-gap. Considering that the incident polarization direction of the $e$ ray is parallel to the incident plane in the Glan-Taylor prism and perpendicular to the incident plane in the Glan-Foucault prism, so we use corresponding reflection indexes for them [6]. Thus the transmitted intensity for the Glan-Taylor prism and for the Glan-Foucault prism is respectively calculated out as

$$I_{fT} = I_0 \cos^2 \phi \left(1 - \frac{tg^2(\theta-\alpha)}{tg^2(\theta+\alpha)}\right)^2 \left(1 + \frac{tg^4(\theta-\alpha)}{tg^4(\theta+\alpha)} + 2\frac{tg^2(\theta-\alpha)}{tg^2(\theta+\alpha)} \cos\frac{4\pi h \cos\alpha}{\lambda}\right), \tag{2}$$

$$I_{fF} = I_0 \cos^2 \phi \left(1 - \frac{\sin^2(\theta-\alpha)}{\sin^2(\theta+\alpha)}\right)^2 \left(1 + \frac{\sin^4(\theta-\alpha)}{\sin^4(\theta+\alpha)} + 2\frac{\sin^2(\theta-\alpha)}{\sin^2(\theta+\alpha)} \cos\frac{4\pi h \cos\alpha}{\lambda}\right), \tag{3}$$

where $\theta$ is the angle of incidence at the crystal-air-gap interface (complementary to the cut angle $S$ of the prism when the light beam normaly incidents on the prism, $\theta = 90° - S$), $\alpha$ is the angle of refraction related to $\theta$ by the law of refraction, $\lambda$ is the wavelength, and $h$ is the thickness of the air-gap, which is uniformly distributed. The right-hand side of Eq. (3) concerned with $\theta$ is expressed as $f(\theta)$. Then the transmitted intensity can be rewritten as

$$I_f = I_0 \cos^2 \phi \times f(\theta) . \qquad (4)$$

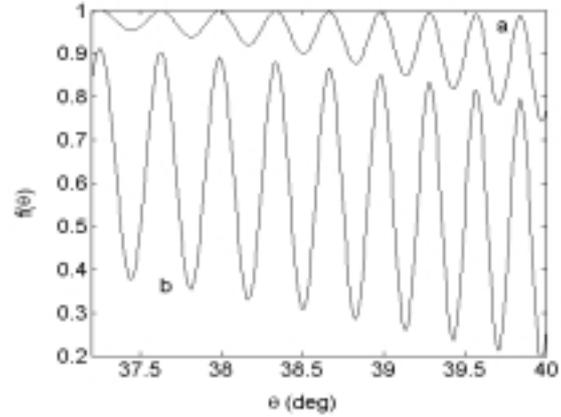

Fig.3. Disturbance factor as an oscillating function of the angle of incidence with $\lambda$=650nm, $n_e$=1.48480, $h$=0.02 mm, (a)Glan-Taylor prism; (b)Glan-Foucault prism.

Figure 3 gives the curves of $f(\theta)$ as an oscillating function of $\theta$. When the incident angle increases, the ampliude and frequency of $f(\theta)$ increase while the average value decreases. In the following, $f(\theta)$ will be named the disturbance factor. The transmitted intensity will be a cosine-squared function for a constant value of $\theta$. However, the prism will be subjected to small vibrations with the rotation of the stepmotor, and as a result the angle of incidence $\theta$ will fluctuate around $\theta_0$ (complementary to $S$) as $\theta = \theta_0 \pm \Delta\theta$ when the stepmotor rotates. The disturbance factor is a high frequency oscillating function as shown in Fig.3, so small disturbances in the angle of incidence will induce big variations of the disturbance factor. Considering the disturbance, the transmitted intensity can be expressed as

$$I_f = I_0 \cos^2 \phi \times f(\theta_0 \pm \Delta\theta). \qquad (5)$$

Since variations of $\theta$ are independent of $\phi$, the disturbance will be stronger where $I_0 \cos^2 \phi$ is greater. As a result, the disturbance is the strongest at the peak of the transmitted intensity, as shown in Fig.2.

Since the amplitude and frequency of the disturbance factor increases as the angle of incidence increases, while the transmittance decreases simultaneously, so a small angle is preferable, however, the beam quality will be affected if the incident angle approaches the angle of total reflection of the ordinary ray ($37.2°$). Considering these two aspects, we had better choose $\theta_0$ as near $38.75°$, which is about the middle angle over the allowable range. It is in accordance with the selection considering the field angle and the extinction ratio in the literature [1]. From Fig.3 we notice that the curve is flattest at the peak, with the weakest disturbance and the highest transmittance. So we'd better choose an angle corresponding to the maximum value of $f(\theta)$ around $38.75°$ as $\theta_0$. As shown in Fig.3, the amplitude of the disturbance factor for the Glan-Foucault prism is much greater than that for the Glan-Taylor prism for a same value of $\theta$, so the disturbance will be much stronger.

The above analysis shows that a proper selection of the value $\theta_0$ will minimize the disturbance. Now we will consider another problem. We may use different light beams with the wavelength varying from about $300$ to $2500$ nm for different experiments. In the meantime the allowed thickness of the air-gap is in a relatively broad range. So how the disturbance is affected by the variation of the wavelength and the thickness of the air-gap deserves careful considerations.

To quantitatively compare the amplitude of the disturbance, we calculate the angles, where the disturbance factor gets its maximum values, then pick out the angle that is nearest to $38.75°$, and take it as the incident angle $\theta_0$. Now we compute $\Delta f(\theta_0, \lambda, h)$ for a same fluctuation amplitude ($0.01°$) of the indecent angle for each group of definite values of $\lambda$ and $h$,

$$\Delta f(\theta_0, \lambda, h) = f(\theta_0, \lambda, h) - f(\theta_0 + 0.01°, \lambda, h). \tag{6}$$

And $\Delta f(\theta_0, \lambda, h)$ properly indicates how great the disturbance of the transmitted intensity is for the same fluctuation of the angle of incidence. The curves of $\Delta f(\theta_0, \lambda, h)$ versus $\lambda$ ($370 \sim 800$ nm) and $h$ ($0.01 \sim 0.03$ mm) are shown in Fig.4, from which we notice that the shorter the wavelength is, the stronger the disturbance will be for a constant thickness of the air-gap. However, the disturbance will decrease as the thickness of the air-gap with the equivalent wavelength decreases. Figure 4 also shows that $\Delta f(\theta_0, \lambda, h)$ for the Glan-Foucault prism is about four times that for the Glan-Taylor prism for the same $\lambda$ and $h$.

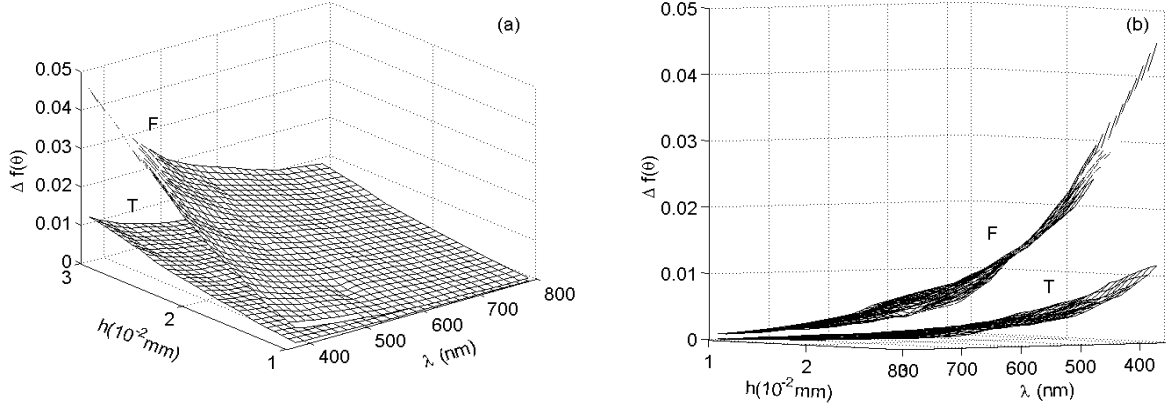

Fig.4. Curves of $\Delta f(\theta_0, \lambda, h)$ versus the wavelenth and the thickness of the air-gap seen from two different look-out angles, where T refers to the Glan-Taylor prism and F refers to the Glan-Foucault prism separately.

Consider two groups of the parameters: $\lambda = 394$ nm, $n_e = 1.49810$, $\theta_0 = 38.76°$ (corresponding to a maximum value of the disturbance factor) and $h = 0.025$ mm; $\lambda = 768$ nm, $n_e = 1.48259$, $\theta_0 = 38.52°$ (corresponding to a maximum value of the disturbance factor) and $h = 0.015$ mm. Substitute these two groups of parameters separately into Eqs. (2) and (3), then draw up the curves of the transmitted intensity vs. $\phi$ for the Glan-Taylor prism and for the Glan-Foucault prism respectively, as shown in Fig.5 (the unit for the vertical coordinates is $I_0$). Comparing Fig.5 (a) and (c) ($\lambda = 394$ nm, $h = 0.025$ mm) with Fig.2, we can see that the theoretical curves are well agreed with the measured ones. Fig.5 (b) and (d) show the disturbance is distinctly reduced with the greater wavelength of $768$ nm and the smaller thickness of the air-gap of $0.015$ mm. To simulate variations of $\theta$ with the rotation of the stepmotor, we have used the random number generator to generate minor variations of angle, which are plus to the value of $\theta_0$ as the angle of incidence $\theta$ ($\theta = \theta_0 \pm \Delta\theta$). It can be seen from Fig.5 that the disturbance of the transmitted intensity for the Glan-Foucault prism is really much stronger than that for the Glan-Taylor prism, at the same time with a lower transmittance, just as what the experiment has exhibited.

There are several feasible methods for reducing the disturbance. Firstly, when designing an air-spaced Glan-type polarizing prism, one should choose the thickness of the air-gap as small as possible in the allowable range [7]. Secondly, for the given wavelength and the definite thickness of the air-gap, the cut angle should be chosen as the complementary angle of the optimal value of $\theta_0$ (at which $f(\theta)$ acquires the maximum value around $38.75°$). Thirdly, one should use the Glan-Taylor prism in place of the Glan-Foucault prism as the polarizer is a better choice. Lastly, take some measures to reduce the vibration of the prism which might weaken the disturbance.

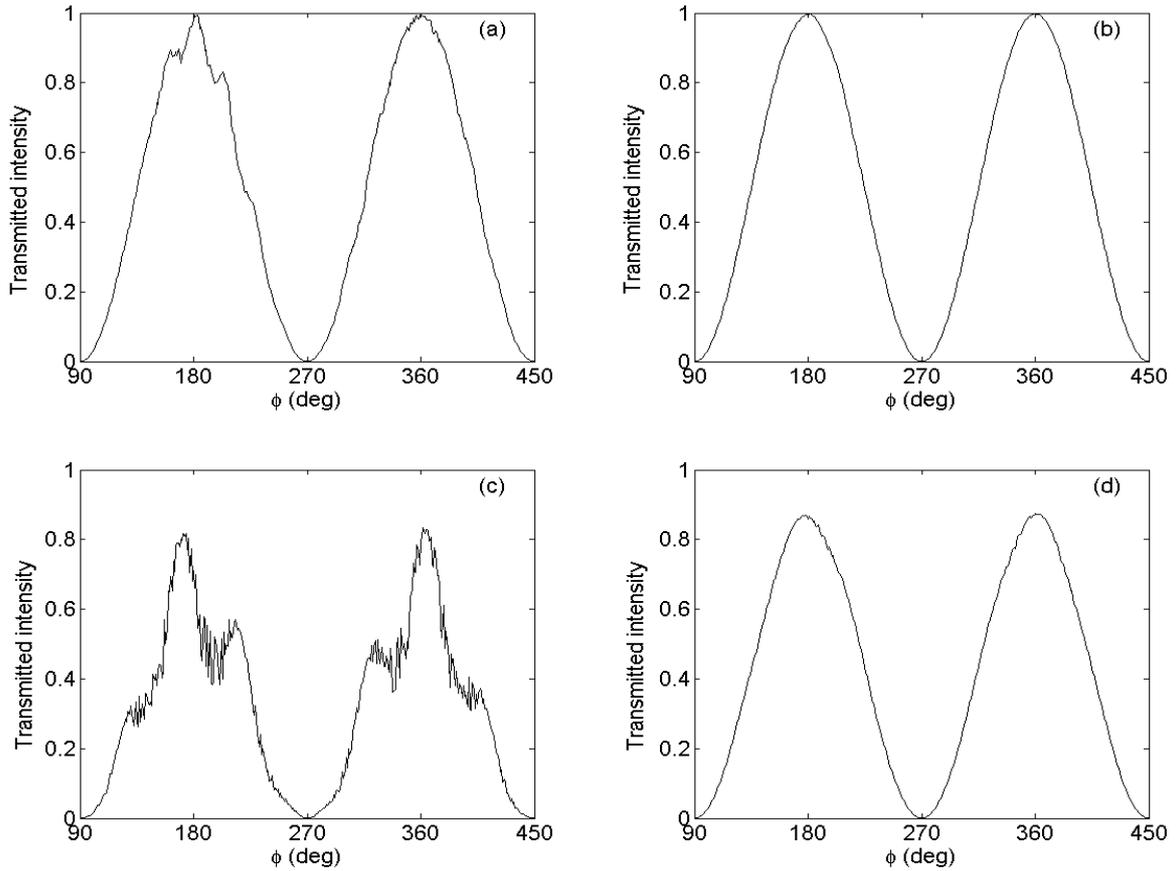

Fig.5. Theoretical curves of the transmitted intensity as a function of $\phi$ with (a)、(c) $\lambda=394$nm, $n_e=1.49810$, $h=0.025$mm, $\theta_0=38.76^o$; (b)、(d) $\lambda=768$nm, $n_e=1.48259$, $h=0.015$mm, $\theta_0=38.52^o$. (a) and (b) are for Glan-Taylor prisms, (c) and (d) are for Glan-Foucault prisms, respectively.

## 4. Conclusions

In conclusion, we have presented, for what is believed to be the first time, the explanation for the disturbance of the transmitted intensity in the air-spaced Glan-type polarizing prisms observed in the experiment. Measures for reducing the disturbance are provided to efficiently improve the quality of the transmitted linearly polarized light. The theories can easily applied to other kinds of prisms, for example, the cemented Glan-type prisms, for which the index of refraction in the air-gap is different from here; and the polarizing beam-splitter prisms, in which the ordinary ray is also transmitted and should be considered too.